\documentclass[twocolumn,showpacs,aps,prd,floatfix]{revtex4}
\usepackage{graphicx}       
\usepackage{dcolumn}           
\usepackage{epsfig}
\usepackage{amsmath}    
\RequirePackage{xspace}

\newcommand{\SLACPubNumber} {13703}         
\newcommand{\LANLNumber} {0907.1743}

\input babarsym

\def\epem  {\ensuremath{e^+e^-}\xspace}
\newcommand\dbline{\noalign{\vskip 0.10truecm\hrule}\noalign{\vskip 2pt}\noalign{\hrule\vskip 0.10truecm}}
\providecommand{\tbline}{\noalign{\vskip 0.05truecm\hrule\vskip0.05truecm}}

\newcommand\etal{{\it et al.}}
\newcommand{\bma}[1]{\boldmath{$#1$}}
\newcommand{\half}{\ensuremath{{1\over2}}}
\newcommand{\pvec}{{\bf p}}
\newcommand{\calB}{\ensuremath{{\cal B}}}
\providecommand{\bfemsix}{${\cal B} (10^{-6})$}

\newcommand{\DE}{\ensuremath{\Delta E}}
\newcommand{\UfourS}{\ensuremath{\Upsilon(4S)}}
\newcommand{\thetaT}{\ensuremath{\theta_{\rm T}}}
\newcommand{\costhr}{\ensuremath{\cos\thetaT}}
\newcommand{\xf}{\ensuremath{{\cal F}}}
\newcommand{\hel}{\ensuremath{{\cal H}}}
\newcommand{\mres}{\ensuremath{m_{\rm res}}}

\newcommand{\signf}{$\cal S$ ($\sigma$)}
\newcommand{\eff}{$\epsilon$ (\%)}

\newcommand{\fetaggetagg}{\ensuremath{\eta_{\gamma\gamma}\eta_{\gamma\gamma}}}
\newcommand{\fetaggetappp}{\ensuremath{\eta_{\gamma\gamma}\eta_{3\pi}}}
\newcommand{\fetapppetappp}{\ensuremath{\eta_{3\pi}\eta_{3\pi}}}

\newcommand{\fetaggphi}{\ensuremath{\eta_{\gamma\gamma}\phi}}
\newcommand{\fetapppphi}{\ensuremath{\eta_{3\pi}\phi}}

\newcommand{\fetaggomega}{\ensuremath{\eta_{\gamma\gamma}\omega}}
\newcommand{\fetapppomega}{\ensuremath{\eta_{3\pi}\omega}}

\newcommand{\fetagghp}{\ensuremath{\eta_{\gamma\gamma} h^+  }}
\newcommand{\fetaggkz}{\ensuremath{\eta_{\gamma\gamma} K^0  }}
\newcommand{\fetapppkz}{\ensuremath{\eta_{3\pi} K^0}}

\newcommand{\etagghp}{\ensuremath{\Bp \ra \fetagghp}}

\newcommand{\fetaggkzs}{\ensuremath{\eta_{\gamma\gamma} \KS  }}
\newcommand{\etaggkzs}{\ensuremath{\Bz \ra \fetaggkzs}}

\newcommand{\fetaggkp}{\ensuremath{\eta_{\gamma\gamma} K^+  }}
\newcommand{\fetapppkp}{\ensuremath{\eta_{3\pi} K^+}}

\newcommand{\fetaggpip}{\ensuremath{\eta_{\gamma\gamma} \pi^+  }}
\newcommand{\fetappppip}{\ensuremath{\eta_{3\pi} \pi^+}}

\newcommand{\fetapeppetapepp}{\ensuremath{\etapr_{\eta\pi\pi}\etapr_{\eta\pi\pi}}}
\newcommand{\fetapeppetaprg}{\ensuremath{\etapr_{\eta\pi\pi}\etapr_{\rho\gamma}}}

\newcommand{\fetapeppkz}{\ensuremath{\etapr_{\eta\pi\pi} K^0}}
\newcommand{\fetaprgkz}{\ensuremath{\etapr_{\rho\gamma} K^0}}

\newcommand{\fetapeppkp}{\ensuremath{\etapr_{\eta\pi\pi} K^+}}
\newcommand{\fetaprgkp}{\ensuremath{\etapr_{\rho\gamma} K^+}}

\newcommand{\fetapepppip}{\ensuremath{\etapr_{\eta\pi\pi} \pi^+}}
\newcommand{\fetaprgpip}{\ensuremath{\etapr_{\rho\gamma} \pi^+}}

\newcommand{\fetapeppphi}{\ensuremath{\etapr_{\eta\pi\pi} \phi}}
\newcommand{\fetaprgphi}{\ensuremath{\etapr_{\rho\gamma} \phi}}

\newcommand{\fetapeppomega}{\ensuremath{\etapr_{\eta\pi\pi} \omega}}
\newcommand{\fetaprgomega}{\ensuremath{\etapr_{\rho\gamma} \omega}}

\newcommand{\etak}{\ensuremath{\eta K}}
\newcommand{\etapK}{\ensuremath{\etapr K}}

\newcommand{\fetapkz}{\ensuremath{\etapr K^0}}

\newcommand{\fphikz}{\ensuremath{\phi K^0}}

\newcommand{\phitoKpKm}{\ensuremath{\phi\ra\Kp\Km}}

\newcommand{\fpizkzs}{\ensuremath{\pi^0 \KS}}
\newcommand{\pizkzs}{\ensuremath{\Bz\ra\fpizkzs}}

\newcommand{\fpizhp}{\ensuremath{\pi^0 h^+}}
\newcommand{\pizhp}{\ensuremath{\Bp\ra\fpizhp}}

\newcommand{\etagg}{\ensuremath{\eta_{\gaga}}}
\newcommand{\etappp}{\ensuremath{\eta_{3\pi}}}
\newcommand{\etatogg}{\ensuremath{\eta\ra\gaga}}
\newcommand{\etatoppp}{\ensuremath{\eta\ra\pi^+\pi^-\pi^0}}

\newcommand{\etaptoepp}{\ensuremath{\etapr\ra\eta\pip\pim}}
\newcommand{\etapepp}{\ensuremath{\etapr_{\eta\pi\pi}}}
\newcommand{\etaprg}{\ensuremath{\etapr_{\rho\gamma}}}
\newcommand{\etaptorg}{\ensuremath{\etapr\ra\rho^0\gamma}}

\newcommand{\fetakz}{\ensuremath{\eta K^0}}

\newcommand{\retakz}{\ensuremath{1.15^{+0.43}_{-0.38}\pm0.09}}

\newcommand{\uletakz}{\ensuremath{1.8}}

\newcommand{\fetaeta}{\ensuremath{\eta\eta}}

\newcommand{\fetaphi}{\ensuremath{\eta\phi}}

\newcommand{\fetaomega}{\ensuremath{\eta\omega}}

\newcommand{\fetapetap}{\ensuremath{\etapr\etapr}}

\newcommand{\fetapphi}{\ensuremath{\etapr\phi}}

\newcommand{\fetapomega}{\ensuremath{\etapr\omega}}
\newcommand{\etapomega}{\ensuremath{\Bz\ra\fetapomega}}

\newcommand{\fetakp}{\ensuremath{\eta K^+}}
\newcommand{\etakp}{\ensuremath{\Bp\ra\fetakp}}

\newcommand{\fetapip}{\ensuremath{\eta \pi^+}}

\newcommand{\fetapkp}{\ensuremath{\etapr K^+}}

\newcommand{\fetappip}{\ensuremath{\etapr \pi^+}}

\newcommand{\BaBarType}      {PUB}  
\newcommand{\BaBarPubYear}    {}
\newcommand{\BaBarPubNumber}  {09/17}
\newcommand{\kp}{\mbox{$K^+$}}
\newcommand{\hp}{\mbox{$h^+$}}

\begin{document}

\begin{flushleft}
~\\
~\\
\end{flushleft}
\begin{flushright}
~\\
~\\
\babar-\BaBarType-\BaBarPubYear-\BaBarPubNumber \\
SLAC-\BaBarType-\SLACPubNumber \\
hep-ex/\LANLNumber\\
\end{flushright}

\title{
 \large  \bf\boldmath $B$ meson decays to
 charmless meson pairs containing $\eta$ or \etapr\ mesons
 
}
%
\author{B.~Aubert}
\author{Y.~Karyotakis}
\author{J.~P.~Lees}
\author{V.~Poireau}
\author{E.~Prencipe}
\author{X.~Prudent}
\author{V.~Tisserand}
\affiliation{Laboratoire d'Annecy-le-Vieux de Physique des Particules (LAPP), Universit\'e de Savoie, CNRS/IN2P3,  F-74941 Annecy-Le-Vieux, France}
\author{J.~Garra~Tico}
\author{E.~Grauges}
\affiliation{Universitat de Barcelona, Facultat de Fisica, Departament ECM, E-08028 Barcelona, Spain }
\author{M.~Martinelli$^{ab}$}
\author{A.~Palano$^{ab}$ }
\author{M.~Pappagallo$^{ab}$ }
\affiliation{INFN Sezione di Bari$^{a}$; Dipartimento di Fisica, Universit\`a di Bari$^{b}$, I-70126 Bari, Italy }
% ...
\collaboration{The \babar\ Collaboration}
\noaffiliation

\begin{abstract}
We present  updated measurements of the branching fractions for \Bz\ meson
decays to \fetakz, \fetaeta, \fetaphi, \fetaomega, \fetapkz,
\fetapetap, \fetapphi, and \fetapomega, and branching fractions and
\CP-violating 
charge asymmetries for \Bp\ decays to  
\fetapip,  \fetakp, \fetappip, and  \fetapkp.
The data represent  the full dataset of $467 \times 10^{6}$ \BB\ pairs
 collected with the 
\babar\ detector at the \pep2 asymmetric-energy \epem  collider at
the SLAC National Accelerator Laboratory. 
Besides large signals for the four charged $B$ decay modes and for
$\Bz\to\etapr K^0$, we find evidence for three \Bz\ decay modes at
greater than $3.0\sigma$ significance.
We find
$\calB(\Bz\to\fetakz) = (1.15^{+0.43}_{-0.38} \pm0.09)\times10^{-6}$, 
$\calB(\Bz\to\fetaomega) = (0.94^{+0.35}_{-0.30}\pm0.09)\times10^{-6}$, and
$\calB(\Bz\to\fetapomega) = (1.01^{+0.46}_{-0.38}\pm0.09)\times10^{-6}$,
where the first (second) uncertainty is statistical (systematic).
For the $\Bp\to\fetakp$ decay mode, we measure  the charge asymmetry 
$\calA_{ch}(\Bp\to\fetakp) = -0.36 \pm 0.11 \pm 0.03$.
\end{abstract}

\pacs{13.25.Hw, 12.15.Hh, 11.30.Er}

\maketitle

Experimental measurements of  branching fractions and \CP-violating
charge asymmetries in rare  $B$ decays play an important role in 
testing the theoretical predictions of the standard
model and its extensions. 
We report the  results of branching fraction measurements for \Bz\
meson decays to \fetakz, \fetaeta, \fetaphi, \fetaomega, \fetapkz,
\fetapetap, \fetapphi, and \fetapomega\ final states and 
of branching fraction and charge asymmetry measurements for \Bp\ decays to
\fetapip, \fetakp, \fetappip, and \fetapkp~\cite{charge}. 
 We search for charge asymmetry by measuring
\begin{equation}
\calA_{ch} \equiv \frac{\Gamma^- - \Gamma^+}{\Gamma^- + \Gamma^+}
\end{equation}
where $\Gamma^{\pm} =\Gamma(B^{\pm} \ra f^{\pm})$ is the decay width
for a given charged final state $f^{\pm}.$ 
 These branching fraction and charge asymmetry  measurements represent an
improvement over  previous results published by \babar~\cite{PreviousBaBar}
and Belle \cite{BelleResults}.

\begin{figure}[!h]
 \begin{center}
 \includegraphics[angle=0,scale=0.60]{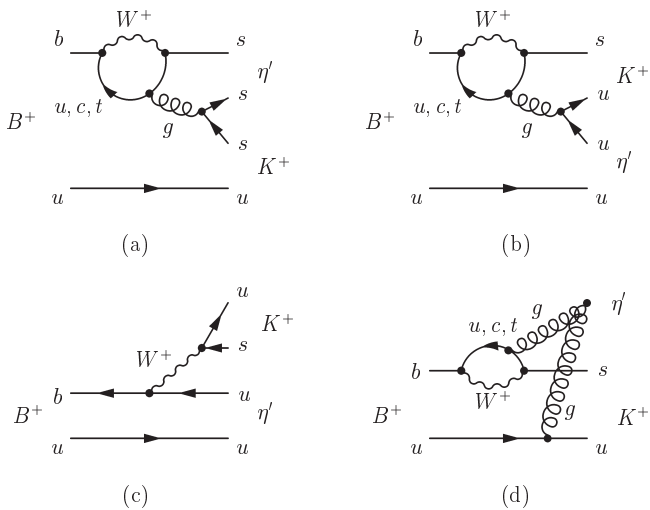}
 \caption{
Examples of Feynman diagrams involved in decays studied in this
paper: (a,b) penguin diagrams, (c) Cabibbo-suppressed tree diagram, (d)
gluonic penguin diagram.
}
 \label{fig:feyn}
 \end{center}
 \end{figure}

The branching fractions and charge asymmetries of the
charmless hadronic $B$ decays are predicted using
approaches based on QCD
factorization \cite{ALI,LEPAGE,BENEKE,SCET} and flavor SU(3) symmetry
\cite{SU3,CHIANGeta,CHIANGVP}.
These $B$ decays proceed through loop (penguin) and suppressed
tree diagram amplitudes, as shown in
Fig.~\ref{fig:feyn}.
The branching fraction and charge  
asymmetry measurements may 
provide sensitivity to the presence of heavy non-standard model 
particles in the loop diagrams~\cite{loops}.   
The measured \etapK\ branching fraction is found to be much larger
than the \etak\ one~\cite{PreviousBaBar,BelleResults}. Many suggestions
have been proposed to explain such a difference, including flavor singlet 
enhancement~\cite{hairpin}, intrinsic charm~\cite{charming}, and
constructively interfering internal penguin
diagrams~\cite{William,Lipkin}. This last approach is supported by
next-to-leading order QCD factorization calculations~\cite{BENEKE}. 

The \CP-violating parameters $S_{\etapr K}$ and $S_{\phi K}$, measured
in the time-dependent analysis of \fetapkz\ and \fphikz\
decays~\cite{TD}, are expected to equal $S_{\ccbar s}\approx
\sin2\beta$, where $S_{\ccbar s}$ is 
measured in the Cabibbo-Kobayashi-Maskawa (CKM) favored $b\to\ccbar s$
decays, if penguin $b\to s$ transitions are dominant. 
However, CKM-suppressed amplitudes and color-suppressed tree diagrams
can introduce additional weak phases whose contributions may not be
negligible~\cite{Gross,BENEKE,london}. As a consequence,
deviations from $\sin2\beta$ may occur even within the standard model.
Rates of the decay modes to \fetaeta,  \fetaphi, \fetapetap, and \fetapphi\
are used in flavor SU(3)-based calculations of the
$|S_{\ccbar s}-S_{f}|$ (with $f=\etapr K,\,\phi K$)~\cite{Gross} bound. This
bound may be improved by more precise measurements of the branching 
fractions of these modes.

The charge asymmetry is expected to be sizable in $\eta\kp$ and
suppressed in $\etapr\kp$ decays~\cite{BENEKE,CHIANGeta,Barshay}.
However, different approaches predict the two asymmetries to have the
same~\cite{CHIANGeta} or opposite~\cite{BENEKE} signs; precise 
measurement of such asymmetries can discriminate between these models.
Furthermore, the charge asymmetries in $\etapr\pip$ and $\eta\pip$ decays are
expected to be sizable~\cite{BENEKE,CHIANGeta}, with model-dependent
predictions for their magnitudes.
      
The results presented here are based on the full dataset  collected
with the \babar\ detector~\cite{BABARNIM}
at the PEP-II asymmetric-energy $e^+e^-$ collider
located at the SLAC National Accelerator Laboratory.  An integrated
luminosity of 426~fb$^{-1}$, corresponding to 
$N_{\BB}= 467 \times 10^6 $  \BB\ pairs, was recorded at 
the $\Upsilon (4S)$ resonance (center-of-mass energy $\sqrt{s}=10.58\
\gev$). A further 44~\invfb was collected approximately 40~\mev\ below
the resonance (off-peak) for the study of the $\epem\to\qqbar$
background, where $q$ is a $u,\, d,\, s$, or $c$ quark. 

Charged particles are detected, and their
momenta measured, by a combination of a vertex tracker, consisting
of five layers of double-sided silicon microstrip detectors, and a
40-layer drift chamber, both operating in the 1.5 T magnetic
field of a superconducting solenoid. We identify photons and electrons 
using a CsI(Tl) electromagnetic calorimeter (EMC).
Further charged-particle identification (PID) is provided by the average energy
loss (\dedx ) measurements in the tracking devices and by the information
provided by an internally reflecting ring-imaging Cherenkov detector
(DIRC) covering the central region. 

We select $\eta$, $\etapr$, $\phi$, $\rho^0$, $\KS$, $\omega$, and $\pi^0$
candidates through the decays \etatogg\ (\etagg), \etatoppp\ (\etappp), 
\etaptoepp\ with \etatogg\ (\etapepp), \etaptorg\ (\etaprg), \phitoKpKm,  
$\rho^0\ra\pipi$, $\KS \ra \pi^+\pi^-$, $\omega \ra \pi^+ \pi^-
\pi^0$,  and $\pi^0 \ra \gamma\gamma$. We do not study the decay
$\Bz\to\etapr\etapr$ with both $\etapr$ mesons decaying to $\rho\gamma$,
because it suffers large backgrounds.
Requirements applied to the photon energy $E_{\gamma}$ and to the
invariant mass of the $B$ daughters are listed in
Table~\ref{tab:cuts}.
\begin{table}[!htb]
\begin{center}
\caption{Selection requirements on the invariant masses of the signal
  resonances and on the laboratory energies of the 
  photons coming from their decays.}
\begin{tabular}{llc}
\dbline
State & Invariant mass (\mevcc) & $E_{\gamma}$ (\mev)\\
\hline
\piz                      & $\phantom{1}120< \phantom{1}
m_{\gaga}\phantom{11}<150 $       & $\!\!\!\!>30$ \\
Prompt \etagg             & $\phantom{1}505 < \phantom{1}
m_{\gaga}\phantom{11}<585$      & $>100$\\ 
Secondary \etagg\         & $\phantom{1}490< \phantom{1}
m_{\gaga}\phantom{11}<600$       & $>50^{\dagger}$\\ 
\etappp\ in \Bp\ decays   & $\phantom{1}534<\phantom{1}
m_{\pi\pi\pi}\phantom{1}< 561$    & -- \\ 
\etappp\ in \Bz\ decays   & $\phantom{1}535<\phantom{1}
m_{\pi\pi\pi}\phantom{1}< 555$    & -- \\ 
\etapepp\ in \Bp\ and & $\phantom{1}945<\phantom{1}
m_{\eta\pi\pi}\phantom{1}<970$ & -- \\ 
$\;\;\Bz\to\etapr\KS$ decays\\
\etapepp\ in other \Bz\ decays  & $\phantom{1}930<\phantom{1}
m_{\eta\pi\pi}\phantom{1}<990$ & -- \\ 
\etaprg\ in \Bp\ decays   & $\phantom{1}930<\phantom{1}
m_{\rho\gamma}\phantom{11}<980$ & $>200$ \\ 
\etaprg\ in \Bz\to\etaprg\KS    & $\phantom{1}930<\phantom{1}
m_{\rho\gamma}\phantom{11}<980$ & $>100$ \\ 
\etaprg\ in other \Bz\ decays  & $\phantom{1}910<\phantom{1}
m_{\rho\gamma}\phantom{11}<990$ & $>200$ \\ 
$\rho^0$                  & $\phantom{1}470<\phantom{1}
m_{\pi\pi}\phantom{11}< 990$      & -- \\ 
$\omega$                  & $\phantom{1}735<\phantom{1}
m_{\pi\pi\pi}\phantom{1}< 825$    & -- \\ 
$\phi$                    & $1012 < m_{K^+K^-}\! < 1026$ & -- \\
\KS                       & $\phantom{1}486 < \phantom{1} m_{\pi\pi} \phantom{11}<
510$ & --\\ 
\dbline
\end{tabular}
\label{tab:cuts}
\vspace*{-0.3cm}
\begin{flushleft}
$^{\dagger}E_{\gamma}>100\mev$ in the
$\Bp\to\etapepp\kp$ and $\Bp\to\etapepp\pip$ decay modes.
\end{flushleft}
\end{center}
\end{table}
The requirements on the $\eta$ and \etapr\ invariant masses depend on
the decay mode.
Branching  
fractions of charged $B$ decays with \etapr\ in the final state and of
$\Bz\to\etapr\KS$ are higher than those of the other neutral $B$
modes. 
In neutral decay modes we apply a tighter requirement on \etappp\
invariant mass in order to prevent possible contamination from \BB\
background. The different requirements on the \etapr\ mass increase the
purity of the charged $B$ and $\etapr\KS$ modes and enhance the
selection efficiency for the other neutral $B$ decay modes.
The energy (momentum) of the \piz\ ($\eta$) candidates is required
to exceed 200~\mev\ (200~\mevc) in the laboratory frame.
The prompt charged tracks in $\Bp\to\etapr\pip$ and secondary charged tracks
in $\eta$, \etapr, and $\omega$ candidates are required to have 
DIRC, \dedx, and EMC signatures consistent with the pion hypothesis.
After selection, we constrain the $\eta$, 
\etapr, and $\pi^0$ masses to their world average values \cite{PDG2008}. 
The prompt charged track in $\Bp\to\etapr\kp$ is required to be 
consistent with the kaon hypothesis.
The signatures for the charged kaons from $\phi$ decays are required
to be inconsistent with hypotheses for electrons, pions and protons.
For the prompt charged track in  \Bp\ decays to \fetakp\ and
\fetapip, we define the variables
${\cal C}_{K}$ and ${\cal C}_\pi$ as
\begin{equation}
{\cal C}_{K,\pi}=\frac{\theta^{meas}_{K,\pi}-\theta^{exp}_{K,\pi}}
{\sigma^{meas}_{K,\pi}}, 
\end{equation}
where $\theta^{meas}_{K,\pi}$ ($\theta^{exp}_{K,\pi}$) is the
measured (expected) DIRC Cherenkov angle and $\sigma^{meas}_{K,\pi}$ is
its uncertainty, for the kaon and pion hypothesis, respectively.
We require \linebreak $-3 < {\cal C}_K < 13$ and $-13 <
{\cal C}_{\pi} < 3$.
For $\KS$ candidates  we require    a
vertex $\chi^2$ probability larger than $0.001$ and a reconstructed 
decay length greater than three times its uncertainty. 

We reconstruct the $B$ meson candidate  by combining the four-momenta of 
the final state particles and imposing a vertex constraint. 
A $B$ meson candidate is kinematically characterized by the
energy-substituted 
mass $m_{ES} = \sqrt{ s/4 - \pvec^{2}_B }$
and energy difference $\DE = E_B-\half\sqrt{s}$, where ($E_B$,$\pvec_B$)
is the $B$-meson four-momentum vector expressed in
the \UfourS\ rest frame.  
For signal events, the \mes\ and \DE\ distributions peak around
$5.28$~\gevcc\ and zero, respectively.
We require $5.25 <\mes <5.29 \gevcc $ and $|\DE|<0.2\gev$ for all decay
modes except $\Bz\to\eta\KS$, where we require $-0.15<\DE<0.2\gev$ in
order to suppress most of the background from radiative $B$ decays.

Backgrounds arise primarily from random combinations of tracks and
neutral clusters in $\epem\ra\qqbar$ continuum events.
We use large samples of Monte Carlo (MC) simulated~\cite{geant}
events and control samples to optimize criteria to suppress the
background.
We reject continuum events by using the angle
\thetaT\ between the thrust axis of the $B$ candidate in the \UfourS\
frame and that of the rest of the event. 
The  thrust axis
of the $B$ candidate is given by the thrust axis of the $B$ decay products.
The distribution of $|\costhr|$ is
sharply peaked near $1.0$ for jet-like \qqbar\ pair events and is
nearly uniform for $\UfourS \rightarrow \BB$ events.  We require 
$|\costhr|<0.9$ (0.85 for \fetaprgpip, 0.8 for \fetaggomega\ and
\fetaprgomega). To discriminate against $\tau$-pair and two-photon backgrounds,
and to better describe the event shape,  we require the
event to contain at least  three charged tracks, or one track beyond
the minimum required for the signal decay topology, whichever
is larger. 

In $\eta\to\gamma\gamma$ ($\phi$) decays, 
we define 
$\mathcal{H_{\eta}}$ ($\mathcal{H_{\phi}}$) as the cosine of the angle between 
the direction of a daughter $\gamma$ ($K$) and the flight direction of the
parent of $\eta$ ($\phi$)  in the $\eta$ ($\phi$) rest frame; for \etaprg,
$\mathcal{H_{\rho}}$ is the cosine of the angle between the direction of 
a daughter pion and the flight direction of the \etapr\ in the
$\rho$ rest frame. For $B$ decays containing an $\omega $ meson in the
final state we define $\mathcal{H_{\omega}}$ as the cosine of the angle
between the $B$ recoil direction and the normal to the plane
defined by the $\omega$ daughters in the $\omega$ rest frame.
We require $|\mathcal{H_{\eta}}|<0.95$ in $\Bz\to\eta\eta$ decay modes.
We reject candidate events if $|\mathcal{H_{\rho}}|>0.9$ ($>0.75$
in the $\Bp\to\etaprg\pip$ decay mode).

For the \etaggkzs\ (\etagghp, $\hp = \kp,\pip$) decay, the main source
of \BB\ background is the \pizkzs\ (\pizhp) decay. To suppress
this background, we search for $\pi^0$ candidates with a photon in
common (overlapping) with the $\eta$ candidate from the reconstructed
signal $B$ candidate. We require the $\pi^0$ mass not to be in the
range (0.117, 0.152)~\gevcc for the \etaggkzs\ decay mode, and
(0.118, 0.150)~\gevcc\ for the $\etagghp$ decay modes.   
Further suppression of this background is obtained with suitable
requirements on $|\mathcal{H_{\eta}}|$ and on the energy of the 
second (non-overlapping with $\eta$) \piz\ photon
($E^{2nd}_{\gamma}$). We optimize these requirements by maximizing
$S/\sqrt{S+B}$, where $S$ ($B$) is the number 
of signal (background) events surviving the selection.
We find the optimal criteria to be $|\mathcal{H_{\eta}}| <
0.966$ and $E^{2nd}_{\gamma}<0.207$~\gev\ for the \etaggkzs\
decay mode, and $|\mathcal{H_{\eta}}| < 0.977$ and
$E^{2nd}_{\gamma}<0.143$~\gev\ for the $\etagghp$ decay modes.

We find  a mean number of $B$ 
 candidates per event in the range 1.0--1.4, depending on the final
 state. Signal events are
 divided into two categories: correctly reconstructed (CR)  signal where all
 candidate particles come from the correct signal $B$, and self
 cross-feed (SCF) signal where at least one candidate particle is
 exchanged with a particle coming from the rest of the
 event. Simulations show that the fraction of SCF 
 candidates is in the range (3--7)\% in charged $B$ decay modes and
 (2--20)\% in neutral $B$ decay modes. If an event has
 multiple $B$ candidates, we select the candidate with the highest $B$
 vertex $\chi^2$ probability, determined from a vertex fit that
 includes both charged and neutral particles~\cite{TreeFitter}. This
 algorithm selects the correct candidate, if present,
 with an efficiency of (91--99)\% and introduces
 negligible bias.   

We obtain yields from unbinned extended maximum-likelihood
(ML) fits.  The main input observables are \DE, \mes, and a
Fisher discriminant \xf\ \cite{Fisher}.
 Where  relevant, the invariant masses \mres\ 
of the intermediate resonances and angular variables \hel\ are used.
The Fisher discriminant \xf\  combines five variables: the angles with
respect to  the beam axis of the $B$ momentum and $B$ thrust axis, the
zeroth and second angular moments $L_{0,2}$  
of the energy flow about the $B$ thrust axis, and the absolute value
of the continuous output of a flavor-tagging algorithm.
The first four variables are evaluated in the \UfourS\ rest frame.
The moments are defined by $ L_r = \sum_s
p_s\times\left|\cos\theta_s\right|^r$, 
where $\theta_s$ is the angle with respect to the $B$ thrust axis of
track or neutral cluster $s$ with momentum $p_s$, and the sum
excludes the $B$ candidate.
Flavor tagging information is derived from an analysis of the decay
products of the non-signal candidate $B$ meson ($B_{\rm tag}$),
using a neural network based technique~\cite{Tagging}.
The output value of the tagging algorithm reflects the different
final states identified in $B_{\rm tag}$ decay.
In particular, the presence of a lepton in the final state usually
results in a large tagging output value, for both \BzBzb\ and \BpBm\
events. Since leptons are not generally present in continuum
background events, the inclusion of the tagging algorithm output in
\xf\ improves its discriminating power between continuum background
and \BB\ events.
The coefficients of \xf\ 
are chosen  to maximize the separation between
the signal and the continuum background. They are determined from
studies of MC signal events and  off-peak data.

The set of probability density 
functions (PDF) used in the
ML fits, specific to each  decay mode, is determined on the basis of studies
with MC samples.
We estimate \BB\ backgrounds using MC  samples   of $B$ decays.
Where needed, we add components to account for \BB\ background events
with a \mes\ or \DE\ distribution that peaks in the signal region and
for background from $B$ meson decays with charmed particles in the
final state.


\begin{table*}[!tph]
\caption{
Fitted signal event  yield and  fit bias in events (ev), detection
efficiency  $\epsilon$, daughter branching fraction product $\prod\calB_i$,
significance $\cal S$, and measured branching
fraction \calB\ with statistical error for each \Bz\  decay mode. For the
combined measurements we give
the significance  (with systematic uncertainties included) and the
branching fraction 
with the statistical and systematic uncertainties (in parentheses the  90\%
CL upper limit). Significances greater than 7 standard deviations
($\sigma$) are omitted. 
}
\label{tab:resultsNeutri}
\begin{tabular}{lccccccc}
\dbline
Mode& \quad Yield (ev) \quad&\quad Fit bias (ev) &\quad \eff \quad &\quad
$\prod\calB_i$ (\%) \quad&\quad \signf \quad &\quad \bfemsix \quad &  \\
\tbline
~~\fetaggkz  &   $21^{+10}_{-9}$ &$0$ &$32.1 $&$13.6 $&$2.5$ &$1.03^{+0.49}_{-0.44}$ \\
~~\fetapppkz  &  $12^{+7}_{-6}$  &$0$ &$20.6 $&$7.9 $ &$2.5$ &$1.56^{+0.92}_{-0.79}$\\
\bma{\fetakz} &          &  & &  &\bma{3.5} &  \bma{\retakz}& \bma{~
  (< \uletakz)} \\
\hline
~~\fetaggetagg &   $13^{+10}_{-9}$ &$+1$ &$23.9$&$15.5$&$1.4$&$0.7^{+0.6}_{-0.5}$ \\
~~\fetaggetappp &  $9^{+6}_{-5}$  &$+1$ &$18.0$&$17.9$&$1.5$&$0.5^{+0.4}_{-0.3}$ \\
~~\fetapppetappp & $0.2^{+2.4}_{-1.7}$  &$-0.1$ &$11.1$&$5.2$
&$0.1$&$0.1^{+0.9}_{-0.6}$ \\
\bma{\fetaeta}&                   &  &
&&\bma{1.9}&\bma{0.5 \pm 0.3\pm 0.1} &\bma{~ (< 1.0)}      \\

\hline
~~\fetaggphi &   $0^{+6}_{-5}$ &$0$ &$29.3$&$19.4$&$0.1$&$0.0 \pm 0.2$ \\
~~\fetapppphi &  $4^{+4}_{-3}$ &$0 $ &$18.3$&$11.2$&$1.9$&$0.4^{+0.4}_{-0.3}$  \\
\bma{\fetaphi} &                   & &&  &\bma{1.4} &\bma{0.2\pm
  0.2\pm 0.1}& \bma{~ (<0.5 )} \\

\hline
~~\fetaggomega &$36^{+13}_{-12}$ &$+3$ &$18.7$&$35.1$&$3.4$&$1.08^{+0.42}_{-0.39}$ \\
~~\fetapppomega &  $8^{+7}_{-5}$ &$+1$ &$13.1$&$20.2$&$1.8$&$0.59^{+0.57}_{-0.40}$ \\
\bma{\fetaomega} & & &&  &\bma{3.7} &\bma{0.94^{+0.35}_{-0.30} \pm 0.09}& \bma{~ (<1.4 )}     \\

\hline
~~\fetapeppkz &  $490^{+25}_{-24}$ &$-2$&$26.6$&$6.1$&$-$&$64.9^{+3.3}_{-3.2}$ \\
~~ \fetaprgkz&   $1003\pm41$ &$+27$&$28.3$&$10.2$&$-$&$72.4 \pm 3.0$ \\
\bma{\fetapkz} & & &&  &\bma{-} &\bma{68.5 \pm2.2 \pm 3.1}&    \\

\hline
~~\fetapeppetapepp &  $1.6^{+2.1}_{-1.1}$ &$0$  &$19.9$&$3.1$&$2.2$&$0.6^{+0.7}_{-0.3}$ \\
~~ \fetapeppetaprg&   $8^{+9}_{-7}$  &$+2$&$19.8$&$10.3$&$0.8$&$0.6^{+0.9}_{-0.7}$ \\
\bma{\fetapetap} &       & &&  &\bma{1.0} &\bma{0.6^{+0.5}_{-0.4}\pm 0.4}& \bma{~ (< 1.7)} \\

\hline
~~\fetapeppphi &  $-2^{+2}_{-1}$ &$0$  &$24.4$&$8.6$&$0.0$&$-0.2^{+0.2}_{-0.1}$ \\
~~\fetaprgphi &   $5^{+8}_{-7}$  &$0$&$23.9$&$14.5$&$0.7$&$0.3^{+0.5}_{-0.4}$ \\
\bma{\fetapphi} &  & &&  &\bma{0.5} &\bma{0.2 \pm0.2\pm 0.3}& \bma{~ (< 1.1)}\\    

\hline
~~\fetapeppomega & $14^{+7}_{-6}$ &$+1$&$17.9$&$15.6$&$3.4$&$1.03^{+0.54}_{-0.46}$ \\
~~\fetaprgomega & $16^{+17}_{-15}$&$-2$&$15.2$&$26.2$&$1.2$&$0.94^{+0.91}_{-0.81}$ \\
\bma{\fetapomega} & & &&  &\bma{3.6} &\bma{1.01^{+0.46}_{-0.38} \pm 0.09}& \bma{~ (< 1.8)} \\    

\dbline
\end{tabular}
\vspace{-5mm}
\end{table*}


The extended likelihood function is
\begin{equation}
{\cal L}= \exp{(-\sum_{j=1}^3 n_j)} \prod_{i=1}^N
\left[\sum_{j=1}^3 n_j  {\cal P}_j ({\bf x}_i)\right],
\end{equation}
where  $N$ is the number of input events,  $n_j$ is the number of
events for hypothesis $j$ ($j=1$ for signal, $j=2$ for continuum
background, and $j=3$ for  \BB\ background), and     ${\cal P}_j ({\bf
  x}_i)$ is the corresponding PDF evaluated   
with the observables ${\bf x}_i$ of the $i^{th}$ event.
In the \etapomega, $\etapr\phi$, and \etaprg$\omega$  decay modes the signal
includes both the CR and SCF signal components with the SCF fraction
fixed to the value estimated  from simulation.
Due to the similar kinematics and branching fractions of the \fetakp\ and
\fetapip\ decay modes, we perform a combined
fit to extract the two signal yields and charge asymmetries.
In this fit we use the ${\cal C}_K$ and ${\cal C}_{\pi}$ variables to
discriminate the mass hypothesis of the prompt track.
Since the correlations among the observables in the data are small,
we assume each  ${\cal P}_j$  to be the product of the PDFs for the
separate variables. Correlations between the $\eta\kp$ and $\eta\pip$
signal yields (charge asymmetries) are below 5\% (7\%).

We determine the PDF functional form and parameters from MC simulation
for the signal and \BB\ backgrounds, and from sideband
data ($5.25 < \mes\ <5.27$ \gevcc; $0.1<|\DE |<0.2$ \gev ) 
for the continuum  background. For $\Bp\to\eta\hp$ 
decay modes, PDF functional form and parameters for the continuum
background are determined using off-peak data.
We parameterize each of
the functions ${\cal P}_{\rm 1}(\mes),\  
{\cal  P}_{\rm 1}(\DE),\ { \cal P}_j(\xf),\ $and the 
peaking components of ${\cal P}_j(\mres)$ with either a symmetric or
a bifurcated
Gaussian, the sum of two symmetric or bifurcated Gaussian shapes, a
bifurcated Gaussian distribution with exponential tails~\cite{Ignoto}
or a Crystal Ball function \cite{crystal}, as required to describe the  
distribution.  Slowly varying distributions ($ \mres$ and $\DE$  
for the continuum background, and angular variables) are represented
by linear or  
quadratic functions.
For the continuum background, the \mes\ distribution
is described by the ARGUS function \cite{Argus}.
Large data control samples of $B$ decays to charmed final states of similar 
topologies are used to verify the simulated resolutions in \mes and \DE.
Where the control samples reveal differences between data and 
 MC samples in mass (energy) resolution, we correct the mean and scale
 the width of the mass (energy) distribution used in the likelihood
 fits.

The validity of the fit procedure and PDF parameterization, including
the effects of unmodeled correlations among observables, is checked
with simulated  
experiments.  
This is done by embedding a number of signal and peaking
\BB\ background events from fully simulated MC samples and by drawing a
number of \qqbar\ and charm \BB\ events from PDFs, according to the
values found in the data. 
In each fit the free parameters are: the yields, the charge asymmetry
for the signal and continuum background, and several parameters describing
the \DE, \mes, and \xf\
distributions of the continuum background. 
A systematic uncertainty due to fixing signal and background parameters
in the fit is accounted.
The charge asymmetry for \BB\ background is fixed to zero in the fit.
A systematic is evaluated to account for this restriction.

\begin{table*}[!ht]
\caption{
Fitted signal event  yield and  fit bias, detection
efficiency  $\epsilon$, daughter branching fraction product $\prod\calB_i$,
measured branching fraction \calB, charge asymmetry $\calA_{ch}$ with
statistical error, and significance $\calS_{\calA}$ of the charge
asymmetry for each charged decay mode.  For the combined measurements
we give the branching fraction,  the charge 
asymmetry and the significance of the charge asymmetry
with the statistical and systematic uncertainties. }
\label{tab:resultsCarichi}
\begin{tabular}{lcccccccc}
\dbline
Mode& \quad Yield (ev) \quad&\quad Fit bias (ev) &\quad \eff \quad &\quad
$\prod\calB_i$ (\%) \quad&\quad \bfemsix & \quad\quad
$\calA_{ch}$ &\quad  $\calS_{\calA}(\sigma)$ \quad \\
\tbline
~~\fetaggpip &$286^{+31}_{-30}$ &$+18$ &$35.1$&$39.3$&$4.16^{+0.48}_{-0.47}$ & $-0.02^{+0.10}_{-0.11}$&$0.4$\\
~~\fetappppip&$95^{+19}_{-18}$ &$+7$ &$23.4$&$22.7$&$3.53^{+0.77}_{-0.73}$ & $+0.06\pm 0.18$ &$0.4$\\
\bma{\fetapip} &          &  & &  &  \bma{4.00 \pm 0.40 \pm 0.24}&
\bma{-0.03\pm 0.09 \pm0.03 }  &\bma{0.3} \\

\hline
~~\fetaggkp &$215^{+31}_{-30}$ &$+21$ &$34.0 $&$39.3$&$3.11^{+0.50}_{-0.48}$ &$-0.37\pm 0.12$ &$3.1$ \\
~~\fetapppkp  &  $69^{+16}_{-15}$  &$+6$  &$22.9$&$22.7$ &$2.60^{+0.66}_{-0.62}$
&$-0.32\pm 0.22$&$1.5$\\
\bma{\fetakp} &          &  & &  &  \bma{2.94^{+0.39}_{-0.34} \pm 0.21 }
&\bma{-0.36 \pm 0.11 \pm 0.03 }  &\bma{3.3}  \\

\hline
~~\fetapepppip &$96^{+20}_{-19}$ &$+1$ &$29.4$&$17.5$&$4.0\pm 0.8$
&$-0.25\pm 0.19$&$1.3$\\ 
~~\fetaprgpip &  $111^{+31}_{-29}$ &$+7$&$25.9$&$29.4$
&$2.9^{+0.9}_{-0.8}$ & $+0.56^{+0.29}_{-0.27}$ &$2.1$\\
\bma{\fetappip} & &  & &  &  \bma{3.5 \pm 0.6 \pm0.2}&\bma{+0.03\pm
  0.17 \pm 0.02 }& \bma{0.2} \\

\hline
~~\fetapeppkp  &   $1601^{+44}_{-43}$ &$-5$ &$28.7$&$17.5$&$68.5
^{+1.9}_{-1.8}$ &$-0.004 \pm 0.027 $ & $0.2$\\
~~\fetaprgkp  &$2991^{+72}_{-71}$&$-10$ &$29.3$&$29.4$&$74.6 \pm 1.8$
&$+0.016\pm 0.023$&$0.7$\\
\bma{\fetapkp} & &  & &  &  \bma{71.5 \pm 1.3 \pm 3.2}&
\bma{+0.008^{+0.017}_{-0.018} \pm 0.009 } &\bma{0.4 }\\

\dbline
\end{tabular}
\vspace{-5mm}
\end{table*}

Table \ref{tab:resultsNeutri} and Table \ref{tab:resultsCarichi} show, 
for \Bz\ and \Bp\ decays, respectively, the measured
yields, fit biases, efficiencies, 
and products of daughter branching fractions for each decay mode. 
The efficiency is calculated as the ratio of the number of signal 
MC events after the event selection to the total generated, and is
corrected for known differences between simulations and data.  
We compute the branching fractions from the fitted  signal event
yields, reconstruction efficiencies, 
daughter branching fractions, and the number of produced $B$ mesons
$N_{\BB}$, assuming equal production  
rates of charged and neutral $B$ pairs from \UfourS\ decays. We
correct the yields for any bias measured with the simulations.
We combine results from different sub-decays by adding the values of
$-2\ln{(\calL/ {\calL_{max}})}$ (parameterized in terms of the branching
fraction or charge asymmetry), where $\calL_{max}$ is the value of
$\cal L$ at its maximum, taking into account  the
correlated and uncorrelated systematic errors. 
We report the branching fractions for the individual decay channels
and their significances ${\cal S}$ in units of standard deviations ($\sigma$). 
For $\Bz\to\etapr\KS$ and all charged decay modes, where the significance of
the branching fraction is always greater than $7\sigma$, the value of
${\cal S}$ is omitted.
For the combined measurements we also report the 90\% confidence
level (CL) upper limits of the branching fraction for the \Bz\ modes where
the significance is less than $5\sigma$.
For charged $B$ decays we give the combined result for the charge
asymmetry $\calA_{ch}$ and its significance $\calS_{\calA}$ in units
of $\sigma$. 

The statistical uncertainty on the signal yield and charge asymmetry
is calculated  
as the change in  the central value when the quantity
$-2\ln{\calL}$ increases by one from its minimum. The
significance is calculated as the square root  of $-2\ln{(\calL_{0}/
  {\calL}_{max})}$, with systematic uncertainties included, where
$\calL_{0}$ is the value of \calL\ for zero signal events or
zero value for the charge asymmetry. 
We determine a Bayesian  90\% CL upper limit on the branching
fraction, assuming a uniform prior probability distribution, by finding
the branching  fraction below which lies 90\% of the total of the
likelihood integral  in the positive branching fraction region.

\begin{figure}[!b]
 \begin{center}
\includegraphics[angle=0,scale=0.45]{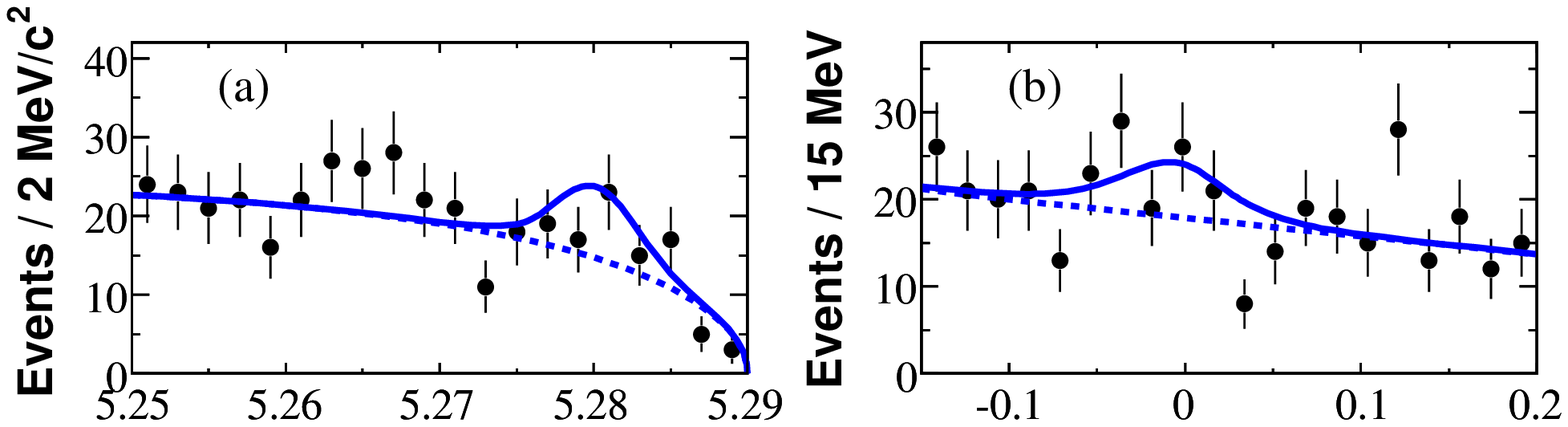}\\
\includegraphics[angle=0,scale=0.45]{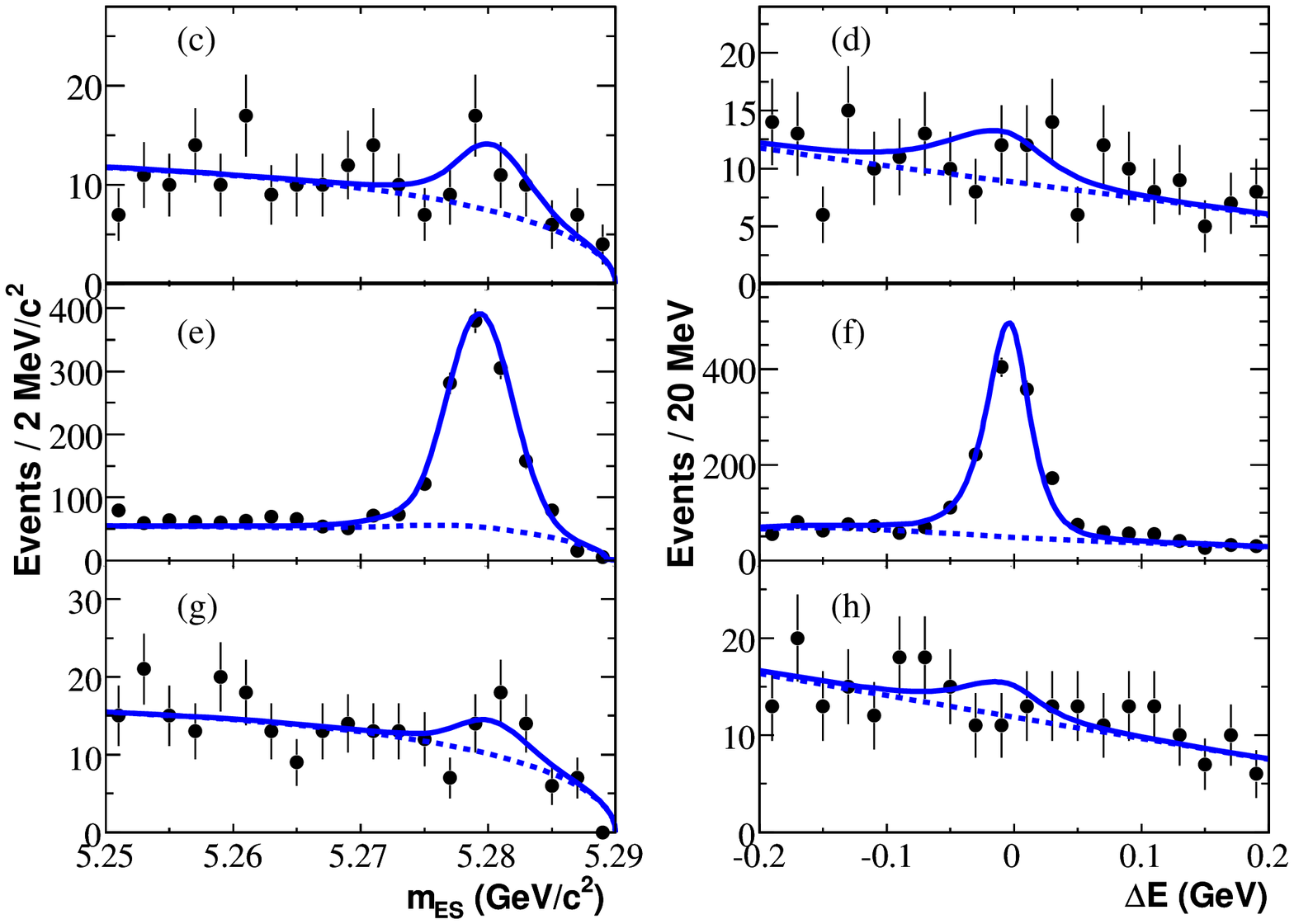}\\
\caption{
 The \Bz\ candidate \mes\ and \DE\ projections for $\eta \KS$ (a, b),
 $\eta\omega$ (c, d), $\etapr \KS$ (e, f), and $\etapr\omega$ (g, h) decays,
 with sub-decays combined.
 Points with errors represent the data, solid curves
 the full fit functions, 
 and dashed curves the background functions.
}
 \label{fig:projMbDENeu}
 \end{center}
 \end{figure}

\begin{figure}[!t]
 \begin{center}
 \includegraphics[angle=0,scale=0.45]{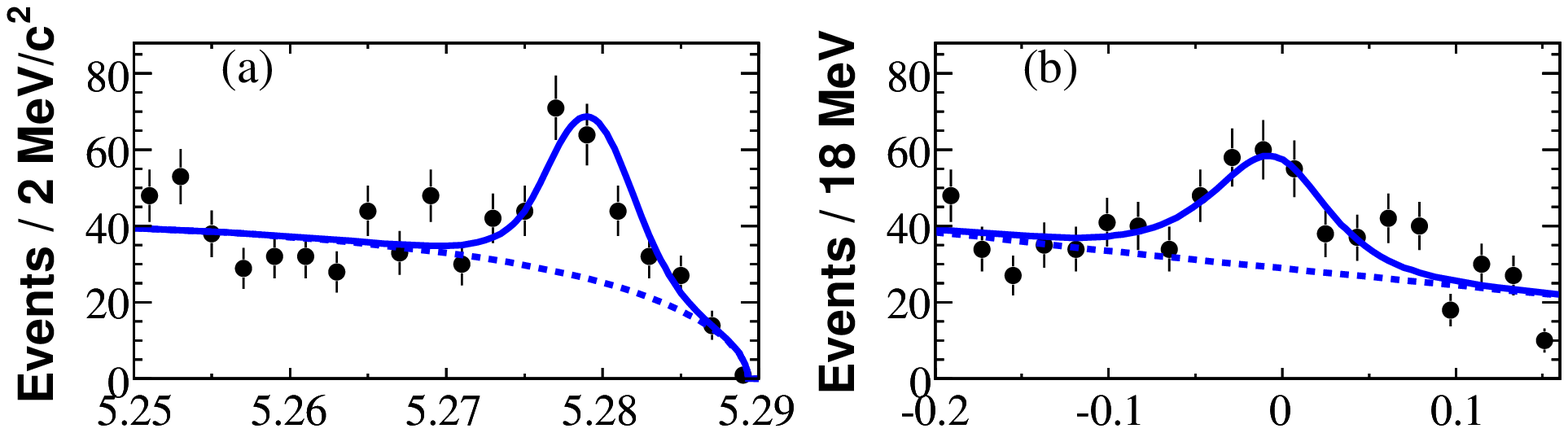}\\
  \vspace*{-.45cm}
   \includegraphics[angle=0,scale=0.45]{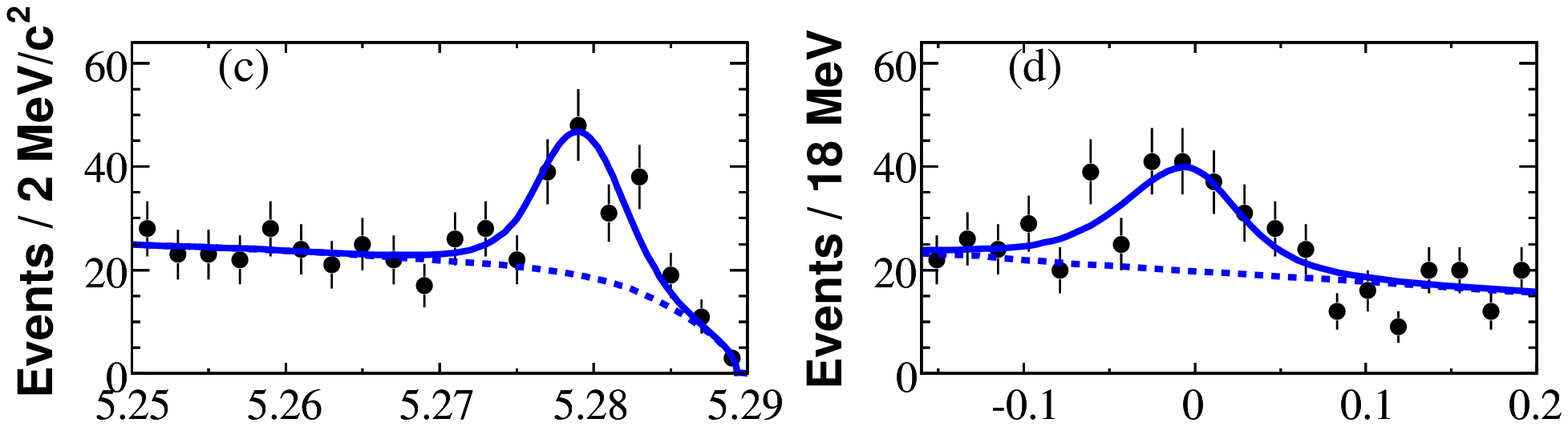}\\
  \vspace*{-.45cm}
  \includegraphics[angle=0,scale=0.45]{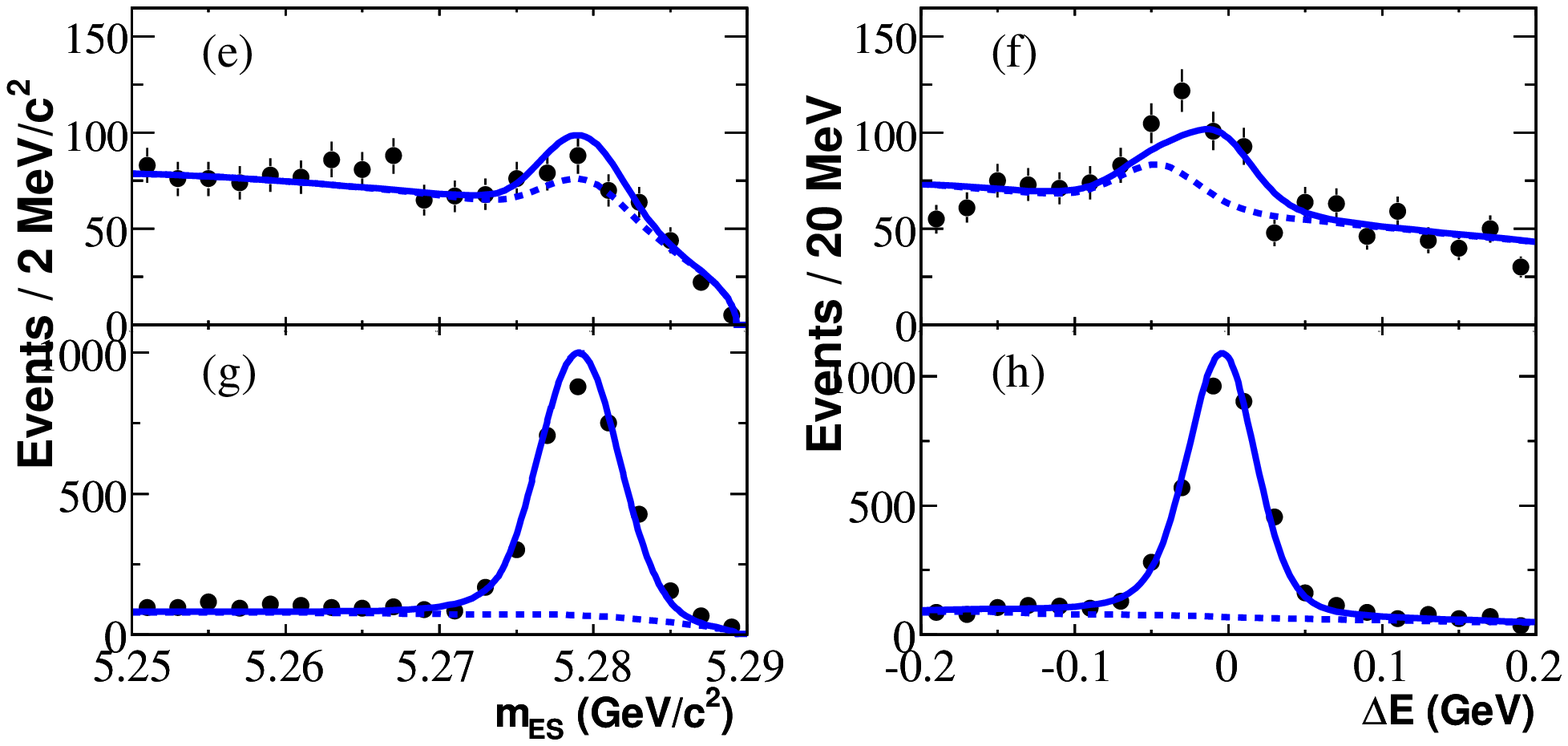}
 \caption{
 The \Bp\ candidate \mes\ and \DE\ projections for \fetapip (a, b),
 \fetakp (c, d), $\etapr\pip$  (e, f), and $\etapr\kp$ (g, h) decays, 
 with sub-decays combined. 
 Points with errors represent the data, solid curves
 the full fit functions, 
 and dashed curves the background functions.
}
 \label{fig:projMbDECar}
 \end{center}
 \end{figure}

Figures\ \ref{fig:projMbDENeu} and~\ref{fig:projMbDECar} show the 
projections onto the \mes\ and \DE\ variables for the four neutral decay
modes that 
have a branching fraction significance greater than $3\sigma$, and for
the four charged decay modes, respectively.
%
%
%
%
For each decay mode we
optimize a requirement on the  probability  ratio    
${\cal P}_{\rm{1}}/({\cal P}_{\rm{1}}+{\cal P}_{\rm{2}} +{\cal P}_{\rm{3}}  )$
in order to enhance the visibility of the signal. The probabilities
${\cal P}_{\rm j}$  are evaluated without using the variable shown.
The points show the data that satisfy such a requirement, while the solid
curves show the total rescaled fit functions.
%
%
%
%
In $\etapr\omega$ decays, a fit performed on $\omega$ mass sidebands
$m_{\pi\pi\pi}<735\mevcc$ or $m_{\pi\pi\pi}>825\mevcc$ shows that
contamination from possible $\Bz\to\etapr\pip\pim\piz$ background is
negligible.  

The main sources of systematic error include ML fit bias (0--14
events) and uncertainties in the PDF 
parameterization (0--12 events). The ML fit bias systematic error is
taken to be half of the bias, 
summed in quadrature with its statistical uncertainty. The
uncertainties related to the PDF 
parameterization are obtained by varying the PDF parameters 
within their errors.
Published world averages
\cite{PDG2008}\ provide the uncertainties of the $B$-daughter branching 
fractions (0--4)\%. These uncertainties are the main contribution to the
systematic errors of the $B\to\etapr K$ decay modes.
The uncertainty on $N_{\BB}$ is  1.1\%. Other  sources of systematic
uncertainty  are track (1\%) and  neutral particle (3--6\%)
reconstruction efficiencies; selection efficiency uncertainties are 1\%
each for the \costhr\  and PID requirements. 
Using large inclusive kaon and $B$ decay samples we estimate a
systematic uncertainty for $\calA_{ch}$ of 0.005 due to the
dependence of the reconstruction efficiency on the charge of the high
momentum $K^{\pm}$. Other sources of systematic uncertainties for
$\calA_{ch}$ are the fit
bias (0--0.02) and the presence of a fit bias in the signal yield
(0.02--0.03).     
The systematic uncertainty due to fixing the value of the charge
asymmetry in \BB\ background components is taken to be the largest deviation
observed when varying this value of $\pm10\%$, and is in range
(0--0.02). 

In summary we present updated measurements 
of branching fractions for eight \Bz\ and four \Bp\  decays to
charmless  meson pairs. The results  shown in Table \ref{tab:resultsNeutri}
and Table \ref{tab:resultsCarichi}  are 
consistent with, but generally more precise than, previous measurements
\cite{PreviousBaBar, BelleResults}  and
supersede our previous ones \cite{PreviousBaBar}.
The branching fraction results are in agreement with predictions within
the theoretical uncertainties  that limit discrimination between different
models~\cite{ALI,LEPAGE,BENEKE,SCET,SU3,CHIANGeta,CHIANGVP}.
We find evidence for three \Bz\ decay modes: \fetakz\
($3.5\sigma$), \fetaomega\ ($3.7\sigma$) and \fetapomega\
($3.6\sigma$).
In the decay mode \etakp\  we find evidence at $3.3\sigma$ for non-zero
charge asymmetry, in agreement 
with theoretical predictions~\cite{BENEKE,CHIANGeta,Barshay}.
Discrimination between QCD factorization~\cite{BENEKE} and
flavor SU(3)~\cite{CHIANGeta} symmetry models, 
based on the relative sign of the charge asymmetry in $\Bp\to\eta\kp$ and
$\Bp\to\etapr\kp$ decays, is limited by the accuracy of the latter
measurement.
The measurement of $\calA_{ch}$ for $\etapr\pip$ shows a slightly better
agreement with the QCD factorization prediction~\cite{BENEKE} than
with the flavor SU(3) symmetry based model~\cite{CHIANGeta}, within
large theoretical and experimental uncertainties.

We are grateful for the 
extraordinary contributions of our \pep2\ colleagues in
achieving the excellent luminosity and machine conditions
that have made this work possible.
The success of this project also relies critically on the 
expertise and dedication of the computing organizations that 
support \babar.
The collaborating institutions wish to thank 
SLAC for its support and the kind hospitality extended to them. 
This work is supported by the
US Department of Energy
and National Science Foundation, the
Natural Sciences and Engineering Research Council (Canada),
the Commissariat \`a l'Energie Atomique and
Institut National de Physique Nucl\'eaire et de Physique des Particules
(France), the
Bundesministerium f\"ur Bildung und Forschung and
Deutsche Forschungsgemeinschaft
(Germany), the
Istituto Nazionale di Fisica Nucleare (Italy),
the Foundation for Fundamental Research on Matter (The Netherlands),
the Research Council of Norway, the
Ministry of Education and Science of the Russian Federation, 
Ministerio de Educaci\'on y Ciencia (Spain), and the
Science and Technology Facilities Council (United Kingdom).
Individuals have received support from 
the Marie-Curie IEF program (European Union) and
the A. P. Sloan Foundation.


\begin{thebibliography}{99}
\bibitem{charge}
Charge conjugation is implied throughout this paper.
\bibitem{Ignoto}
We use the function $f(x)=\exp \left(
\frac{-(x-\mu)^2}{2\sigma^2_{L,R} + \alpha_{L,R} (x-\mu)^2} \right)$,
where $\mu$ is the peak position of the distribution, $\sigma_{L,R}$
are the left, right widths, respectively, and $\alpha_{L,R}$ are the
left, right tail parameters.
\end{thebibliography}
\end{document}